\documentstyle[aps,prl,amssymb]{revtex}
\input{tcilatex}

\begin{document}
\title{The superconducting gap $\Delta _{c}$, the pseudogap $\Delta _{p}$ and pair
fluctuations above $T_{c}$ in overdoped Y$_{1-x}$Ca$_{x}$Ba$_{2}$Cu$_{3}$O$%
_{7-\delta }$ from femtosecond time-domain spectroscopy.}
\author{J.Demsar, B.Podobnik, V.V.Kabanov and D.Mihailovic}
\address{Jozef Stefan Institute, Jamova 39, 1001 Ljubljana, Slovenia}
\author{Th.Wolf}
\address{Forschungszentrum Karlsruhe, Institut fur Technische Physik, D-76021,\\
Karlsruhe, Germany}
\maketitle

\begin{abstract}
The low-energy electronic excitation spectrum and gap structure in optimally
doped and overdoped Y$_{1-x}$Ca$_{x}$Ba$_{2}$Cu$_{3}$O$_{7-\delta }$ single
crystals are investigated by real-time measurements of the quasiparticle
relaxation dynamics with femtosecond optical spectroscopy. From the
amplitude of the photoinduced reflectivity as a function of time,
temperature and doping $x$ we find clear evidence for the coexistence of two
distinct gaps in the entire overdoped phase. One is a temperature-{\em %
independent }''pseudogap'' $\Delta _{p}$ and the other is a $T${\em %
-dependent} collective gap $\Delta _{c}(T)$ which has a BCS-like $T$%
-dependence closing {\em at }$T_{c}$. From quasiparticle relaxation time
measurements above $T_{c}$ we ascertain that fluctuations associated with
the collective gap $\Delta _{c}(T)$ are limited to a few K, consistent with
time-dependent Ginzburg-Landau theory and are distinct from the pseudogap
whose presence is apparent well above $T_{c}$ for all $x.$
\end{abstract}


The phase diagram of high-temperature superconducting copper oxides is
commonly thought to describe the evolution from a Mott-Hubbard insulator at
one end to a more conventional metal on the other. However, the cross-over
from localised states in the ground state on one side to itinerant electrons
on the other is a general and still open problem. Recently, theoretical
calculations\cite{Emery} have led to suggestions that this cross-over may
proceed via a spontaneous self-organisation into an inhomogeneous state with
short range charge (and spin) ordering, rather than through a continuous
transition. Supporting this view, on the experimental side\cite{Review}
there has been mounting evidence in the last few years for an inhomogeneous
charge distribution in the Cu-O planes. The two relevant energy scales in
this state are {\it (i)} the critical temperature $T_{c}$, at which
macroscopic phase coherence is established between hole pairs,
experimentally exhibiting the Meissner state with zero resistivity, and {\it %
(ii)} one or more so-called pseudogaps with energy $\Delta _{p}>kT_{c},$
which manifest themselves through an apparent reduction in the density of
states (DOS) above $T_{c}$. \ To determine the temperature and doping
dependence of all the energy gap(s) in such a multi-component system from
frequency-domain spectroscopy (FDS)\ means extracting gap information by
deconvolution of all the different spectral components. This inevitably
leads to ambiguity in the interpretation of the data, as highlighted by the
numerous controversies regarding the interpretation of infrared, Raman and
photoemission spectra amongst others.

As an experimental alternative, {\em time-domain }spectroscopy (TDS) can
distinguish between different excitations by their different relaxation
dynamics, potentially giving new and complementary information on the
low-lying electronic structure\cite{Stevens,Kabanov}. For example, from the
temperature dependence of the quasiparticle (QP) dynamics in YBa$_{2}$Cu$%
_{3} $O$_{7-\delta }$ (YBCO) it was recently shown that a $T$-independent
charge gap $\Delta _{p}(x)$ is dominant in the underdoped state, whose
magnitude is inversely proportional to doping as $\Delta _{p}(x)\propto 1/x$%
\cite{Kabanov} in good agreement with other measurements on cuprates\cite
{Muller}. Near optimum doping however, the same measurements show a dominant 
$T$-dependent gap $\Delta _{c}(T)$ which closes at{\em \ }$T_{c}$. However,
the fundamental question of how the two gaps evolve near optimum doping and
into the overdoped state still needs to be answered experimentally. In this
letter we report the first direct and detailed TDS measurements of the
evolution of the QP recombination dynamics from optimum doping to the
overdoped region of the phase diagram of a high-$T_{c}$ superconductor Y$%
_{1-x}$Ca$_{x}$Ba$_{2}$Cu$_{3}$O$_{7-\delta }$ as a function of $T$ and $x,$
paying particular attention to the multicomponent response and the evolution
of the ''pseudogap'' $\Delta _{p}(x)$ and the collective gap $\Delta _{c}(T)$
with $x$. The time-domain information on the relaxation of the QP\ density
above $T_{c}$ also enables us to clarify the connection between the
pseudogap and pair fluctuations in the overdoped region.

In time-resolved pump-probe experiments, a laser pulse first excites
electron-hole pairs which then rapidly relax to states close to the Fermi
energy by $e-e$ and $e-ph$ scattering creating a non-equilibrium QP
population. This initial avalanche QP multiplication process occurs within $%
\tau _{m}\sim $100 fs\cite{Kabanov,Brorson}. The presence of a gap near $%
E_{F}$ causes a bottleneck in the final stage of relaxation, so that
carriers accumulate in QP\ states above the gap, giving rise to a transient
change in absorbance or reflectivity which is detected by a second probe
laser pulse. A detailed description of the experimental technique and theory
of femtosecond time-resolved QP spectroscopy can be found elsewhere\cite
{ACS98,Kabanov}. We used light pulses from a Ti:Sapphire laser producing $%
\tau _{L}\lesssim 80$ fs pulses at 800 nm (approx.1.5 eV) for both the pump
and the probe. The photoinduced\ (PI) change in reflectivity ${\sl \Delta }%
{\cal R}/{\cal R}$ was measured using a photodiode and lock-in detection.
The pump laser power was 
%TCIMACRO{\TEXTsymbol{<} }%
%BeginExpansion
\mbox{$<$}%
%EndExpansion
10 mW, exciting approximately 10$^{19}-$ 10$^{20}$ carriers per cm$^{3},$
and the pump/probe intensity ratio was $\sim $100. The steady-state heating
effect was accounted for as described in Ref.\cite{ACS98}, giving an
uncertainty in sample temperature of $\pm 2$K. The experiments were
performed on four Y$_{1-x}$Ca$_{x}$Ba$_{2}$Cu$_{3}$O$_{7-\delta }$ single
crystals with $x=0,$ $0.016,$ $0.101$ and 0.132 and $T_{c}$s of 93K, 89.5K,
83 K and 75 K respectively, grown by the self flux method in Y- or Ca
stabilized ZrO$_{2}$ crucibles. The Ca-content was determined by EDX and
neutron diffraction analysis. The oxygen content $\delta $ was adjusted by
heat treatment and adjustment of oxygen pressure to give $\delta =6.94$ for $%
x$=0, and 6.986, 6.943 and 6.928 for $x=0.016,$ $0.101$ and 0.132
respectively. $T_{c}$ was measured by {\sl dc} magnetization for each $x$
and $\delta $ as shown in the insert to Fig. 1b).

The time-evolution of the PI reflection, ${\sl \Delta }{\cal R}/{\cal R}$,
is shown for $x=0$ and 0.132 at a few temperatures in Figures 1a) and b).
Above $T_{c}$, a single exponential gives a very good fit to the data with a
relaxation time of $\tau _{B}\sim 0.5$ ps. Importantly, we note that beyond
3 ps ($\simeq $6$\tau _{B}$) the signal has decayed to nearly a constant
value, indicating that {\it no other} relaxation process is present on this
timescale\cite{remark1}. This is true for all $0<x<0.132$. This is also
evident from the logarithmic plots in Fig. 1 c) and d). {\em Below }$T_{c}$
however, the logarithmic plots of ${\sl \Delta }{\cal R}/{\cal R}$ shown in
Figs. 1c) and 1d) reveal a\ clear break in the slope near $t=3$ ps,
indicating the presence of two distinct relaxation times, one with $\tau
_{B}\approx 0.5$ ps and the other with $\tau _{A}\approx 3$ ps. This clearly
implies that a two-component fit to the data is necessary for an accurate
description. We therefore model the response as a sum of two components,
each given by the solution of $dr/dt=-r/\tau +G(t),$ where $r=$ ${\sl \Delta 
}{\cal R}/{\cal R}$ and $G(t)$ is the excitation temporal profile
approximated by $G(t)=G_{0}\exp -2[t/\tau _{m}]^{2}$. For $t>200$ fs$,$ we
can simplify the solution to ${\sl \Delta }{\cal R}/{\cal R}\left( t\right)
=A(T)\exp (-t/\tau _{A})+B(T)\exp (-t/\tau _{B}),$ where both amplitudes $%
A(T)$ and $B(T)$ are $T$-dependent and $A(T)=0$ for $T>T_{c}$.

In Fig. 2 we have plotted the relaxation times $\tau _{A}$ and $\tau _{B}$
as a function of temperature for different $x$. \ The common feature for all 
$x$ is the divergence of $\tau _{A}$ just below $T_{c}$ similar to that
reported previously near optimum doping\cite{Kabanov,Han}. In contrast, $%
\tau _{B}$ is found to be completely $T$-independent, as previously observed
in underdoped YBa$_{2}$Cu$_{3}$O$_{7-\delta }$\cite{Kabanov,JPCS98}.

Below $T_{c}$ the QP\ recombination time of the superconductor with a gap $%
\Delta (T)$ can be expressed as\cite{Kabanov}:

\[
\tau =\frac{\hbar \omega ^{2}\ln \left\{ (\frac{{\cal E}_{I}}{2N(0)[{\bf %
\Delta }(0)]^{2}}+e^{-{\bf \Delta }(T)/k_{B}T})^{-1}\right\} }{12\Gamma
_{\omega }[{\bf \Delta }(T)]^{2}}
\]
where $\omega $ is a typical phonon frequency, $\Gamma _{\omega }$ is a
characteristic anharmonic phonon linewidth, $N(0)$ is the DOS and ${\bf %
\Delta }(0)$ is the gap at zero temperature. The important feature of Eq.
(1) is that near $T_{c}$, we can expand the formula for small $\Delta
(T)\rightarrow 0,$ giving $\tau $ $\propto 1/\Delta (T)$\cite{Kabanov},
which means that $\tau $ must diverge at $T_{c}$. On the other hand, if $%
\tau $ is constant, this implies that $\Delta $ is $T$-independent. To model
the divergent signal (component $A$) below $T_{c}$, we substitute $N(0)=5$ eV%
$^{-1}$cell$^{-1}$spin$^{-1}$, $\Gamma _{\omega }=10$ cm$^{-1}$ \cite
{Mihailovic} and $\omega =400$ cm$^{-1}$, with - for simplicity - a BCS\
functional form for $\Delta (T)=\Delta _{c}(T)$ and $\Delta _{c}(0)=4kT_{c}$%
. The result is shown by the solid curves in Fig 2. The clear divergence of $%
\tau $ at $T_{c}$ is evidence for the existence of a collective gap in the
entire overdoped region. On the other hand, the simultaneous presence of a $T
$-independent $\tau _{B}$ indicates the {\em co-existence }of a $T$%
-independent gap $\Delta _{p}$ also over the whole overdoped region.

To obtain more quantitative information on $\Delta _{c}(T)$ and $\Delta _{p}$
we analyse the temperature-dependence of $\left| {\sl \Delta }{\cal R}/{\cal %
R}\right| $ plotted in Fig.3 as a function of $T$. Qualitatively similar
behavior is observed for all $x$: at low $T,$ $\left| {\sl \Delta }{\cal R}/%
{\cal R}\right| $ is nearly constant exhibiting a slight upturn near 0.7 $%
T_{c}$ and then a rapid drop to approximately 30\% of maximum amplitude just
below $T_{c}$. Close to $T_{c}$, there is a clear break in the response and $%
\left| {\sl \Delta }{\cal R}/{\cal R}\right| $ reverts to a much slower
asymptotic temperature dependence above $T_{c}$ extending to 150 K or more.

In the limit of small photoexcited carrier density, we can assume that all
possible contributions to ${\sl \Delta }{\cal R}/{\cal R}$ - arising from
excited state absorption and photoinduced band-gap changes for example - are
linear in the photoexcited carrier density. So, for a $T$-dependent gap $%
\Delta _{c}(T),$ the temperature dependence of the amplitude of the
photoinduced reflectivity ${\sl \Delta }{\cal R}/{\cal R}$ is given by\cite
{Kabanov}:

\[
A(T)=\frac{{\cal E}_{I}/({\bf \Delta }_{c}(T)+k_{B}T/2)}{1+\frac{2\nu }{%
N(0)\hbar \Omega _{c}}\sqrt{\frac{2k_{B}T}{\pi {\bf \Delta }_{c}(T)}}e^{-%
{\bf \Delta }_{c}(T)/k_{B}T}}
\]
where ${\cal E}_{I}$ is the incident energy density per unit cell of the
pump pulse, $\nu $ is the number of phonon modes interacting with the QPs, $%
N(0)\,$is the DOS and $\Omega _{c}\ $is a typical phonon cutoff frequency. A
similar expression gives the amplitude for a $T$-independent gap $\Delta
_{p}:$ 
\[
B(T)=\frac{{\cal E}_{I}/{\bf \Delta }_{p}}{1+\frac{2\nu }{N(0)\hbar \Omega
_{c}}e^{-{\bf \Delta }_{p}/k_{B}T}}.
\]

The two expressions predict qualitatively different $T$-dependence for $%
\left| {\sl \Delta }{\cal R}/{\cal R}\right| $. As $T_{c}$ is approached
from below, Eq. 2 predicts that $A(T)\rightarrow 0$ as $\Delta
(T)\rightarrow 0$. In contrast, Eq. (3) predicts an asymptotic (exponential)
fall of the amplitude at high $T$. Moreover, Eq. 2 predicts a slight maximum
at $T/T_{c}\approx 0.7,$ which is not present for the case of a $T$%
-independent gap (Eq.(3)). These differences between the two predictions
allow us to unambiguously identify the temperature dependence of the QP
gaps, and determine their magnitude. Using $\nu =18,$ $\Omega _{c}=0.1$ eV
and $N(0)=5$eV$^{-1}$cell$^{-1}$spin$^{-1}$ as before, fits to the
temperature-dependence of $\left| {\sl \Delta }{\cal R}/{\cal R}\right| $
with the sum of (2) and (3) are plotted in Fig.3. The values of $\Delta
_{c}(0)$ and $\Delta _{p}$ are also shown in each case. It is evident from
the plots that the total amplitude $\left| {\sl \Delta }{\cal R}/{\cal R}%
\right| $ can only be described accurately by a two component fit and cannot
be described by either component separately. The gap ratios obtained from
the fits are $\Delta _{c}/k_{B}T_{c}\approx 5\pm 0.5$, depending slightly on 
$x$. $\Delta _{p}$ and $\Delta _{c}(0)$ from the fits of the $T$-dependences
of $\left| {\sl \Delta }{\cal R}/{\cal R}\right| $ as a function of doping
are shown in Fig. 4. (The data on $\Delta _{p}$ for underdoped YBCO\cite
{Kabanov} have also been included for completeness.). Remarkably, in the
cross-over region the two gaps converge $\Delta _{p}\rightarrow \Delta
_{c}(0)$, but they {\em remain clearly distinct}, as indicated by the
2-component decay in Figs. 1c) and d), as well as in Fig.2a)-d) and in the $%
T $-dependence analysis of $\left| {\sl \Delta }{\cal R}/{\cal R}\right| $
(Fig.3).

Turning our attention to\ the relaxation dynamics of the order parameter 
{\it above }$T_{c}$, if we assume that Ginzburg-Landau (GL) theory can be
applied to the collective state which exhibits a $T$-dependent gap $\Delta
_{c}(T),$ then the only contribution relevant to the present experiments is
from non-equilibrium pair density fluctuations. Time-dependent GL (TDGL)\
theory\cite{Gorkov} predicts the relaxation time for the amplitude of these
fluctuations above $T_{c}$ to be $\tau _{GL}=\pi \hbar \lbrack
8k(T-T_{c})]^{-1}\simeq 3.0/(T-T_{c})$ psK. Plotting this (parameterless)
expression for $\tau _{GL}$ as a function of $T$ above $T_{c}$ in Fig.2, we
see that $\tau _{GL}$ drops to zero within a few Kelvin of $T_{c},$
consistent with the data on $\tau _{A}$. Any QP density extending {\em %
significantly above} that predicted by TDGL theory would be clearly evident
in the data above $T_{c}$, but it is not. We conclude that pair fluctuations
associated with the{\em \ collective phase }are consistent with TDGL theory
and are quite unrelated to the pseudogap behavior.

To put the present results in the context of other spectroscopy experiments,
the data on $\tau _{A}$ and $\tau _{B}$ suggest that above $T_{c}$ we should
expect a broad peak (gap) at $\Delta _{p}$ of width $\Delta E_{A}\gtrsim
h/(\pi c\tau _{A})(\simeq 4$ meV). Below $T_{c}$ in addition to this peak, a
narrower QP peak should be present with $\Delta _{c}\simeq 5kT_{c}$, with $%
\Delta E_{B}\gtrsim 0.5$ meV. The simultaneous presence of two gaps have
been previously reported in tunneling spectra\cite{Deutscher} and microwave
experiments\cite{Sridhar} on optimally doped YBCO. Comparing our $\Delta
_{p}(x)$ with the gap from Giever (single-particle) tunneling on YBCO, we
find very good agreement (see Fig. 4)\cite{Deutscher}. Extending the
discussion to other cuprates, two-component behavior was reported in La$%
_{2-x}$Sr$_{x}$CuO$_{4}$ over a large portion of the phase diagram, although
so far only a $T$-independent pseudogap was observed. A $T$-independent
pseudogap has also been reported in overdoped Bi$_{2}$Sr$_{2}($Ca,Y)Cu$_{2}$O%
$_{8}$ (BISCO)\cite{Hackl,Fisher,Deutscher}. The co-existence of $\Delta
_{p} $ and $\Delta _{c}(T)$ is also consistent with the $T$-dependence of
the photoemission lineshapes in BISCO\cite{Norman}.

Speculating on the origins of the two-gap behavior in the spatially
inhomogeneous phase picture\cite{Review}, it is natural to associate the
behavior of $\Delta _{c}(T)$ with high carrier density areas, where the gap
in the QP spectrum is formed as a collective effect. Pairing there occurs
simultaneously with macroscopic phase coherence at $T_{c}$ and the
fluctuation region above $T_{c}$ is small, consistent with TDGL theory. On
the other hand, in low-density regions, where no collective effects are
present, $\Delta _{p}$ signifies the individual pair binding energy. The
pseudogap behavior arises due to the fluctuating presence of pairs in the
ground state, whose density is determined by thermal occupancy. However, at $%
T_{c}$ macroscopic phase coherence is established across both regions into a
common superconducting state.

To conclude, the TDS\ experiments on the time- and temperature dependence of
the photoinduced QP response give a self-consistent and systematic picture
of the low-energy charge excitation spectrum of Y$_{1-x}$Ca$_{x}$Ba$_{2}$Cu$%
_{3}$O$_{7-\delta }.$ They suggest that the cross-over from the underdoped
to the overdoped region of the phase diagram occurs via a 2-component
inhomogeneous state with two coexisting gaps, one $T$-independent and one
with a BCS-like $T$-dependence. According to present TDS experiments the
mixed gap region is present over most of the overdoped and optimally doped
phase. To what extent the two gap behavior is universal in the cuprates
remains to be shown.

We would like to thank Airton A.Martin for kindly processing and
characterising the samples used in this investigation.

\thinspace Figure 1. The photoinduced reflection ${\sl \Delta }{\cal R}/%
{\cal R}$ from Y$_{1-x}$Ca$_{x}$Ba$_{2}$Cu$_{3}$O$_{7-\delta }$ above and
below $T_{c}$ as a function of time a)\ for $x=0$ ($T_{c}$=93 K) and b) $%
x=0.132$ ($T_{c}$=75 K) at different temperatures. \ A 2-exponential fit is
made below $T_{c}$ and a single exponential fit above $T_{c}$. In c) and d)
the same data for x=0 and x=0.132 respectively are presented on a
logarithmic scale, clearly showing the single-exponential decay above $T_{c}$
and the 2-exponential decay below $T_{c}$ with a break near 3 ps. The insert
to b) shows the dc magnetization curves for the four samples.

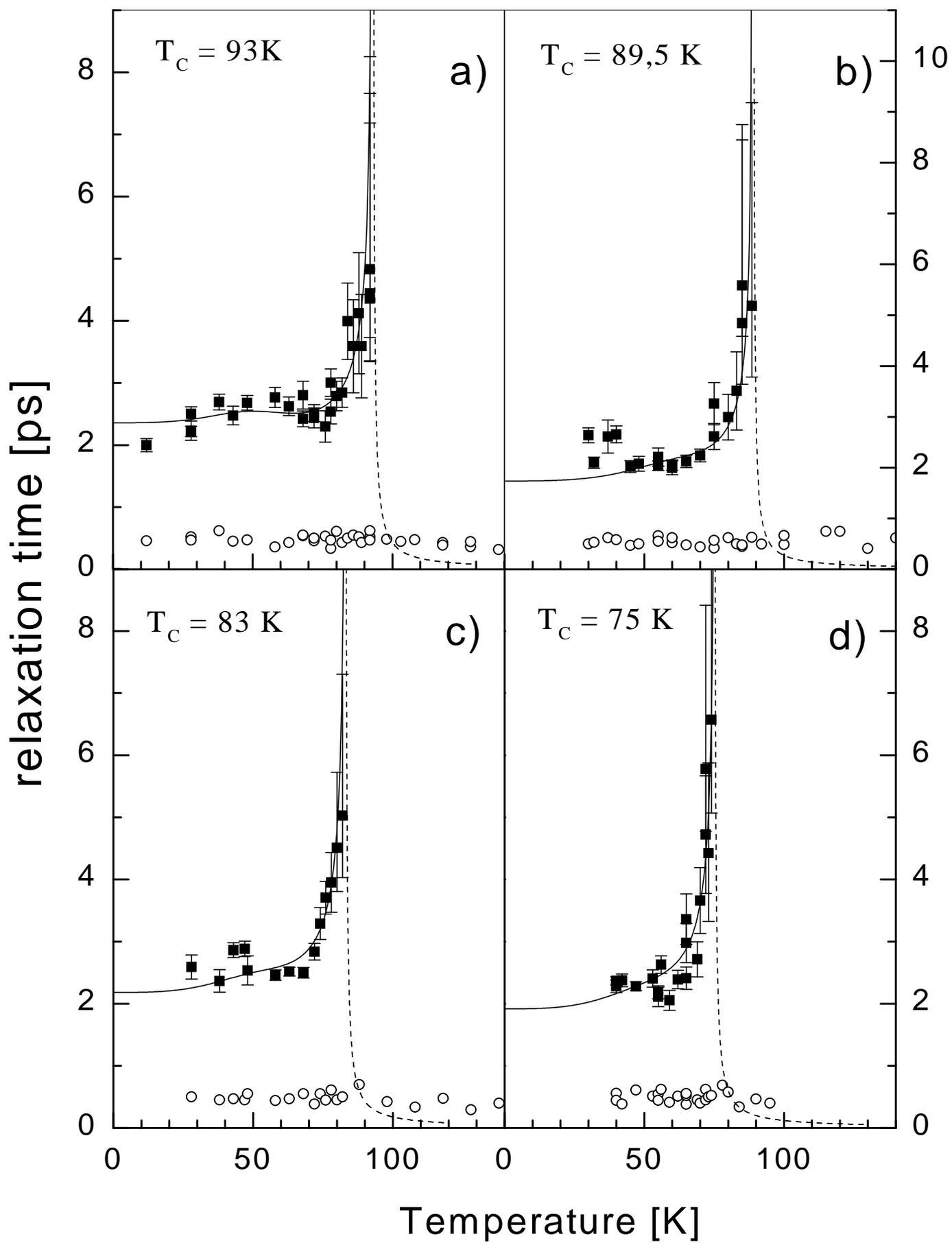
Figure 2. The relaxation times $\tau _{A}$ (squares) and $\tau _{B}$ (open
circles) as a function of $T$ for Y$_{1-x}$Ca$_{x}$Ba$_{2}$Cu$_{3}$O$%
_{7-\delta }$ with a) $x$=0, b) $x$=0.016, c) $x$=0.101 and d) $x$=0.132.
The solid line is the prediction of the relaxation time below $T_{c}$ given
by Eq.(1). The dashed line describes the expected QP\ relaxation time above $%
T_{c}$ given by $\tau _{GL}=\pi \hbar \lbrack 8k(T-T_{c})]^{-1}.$

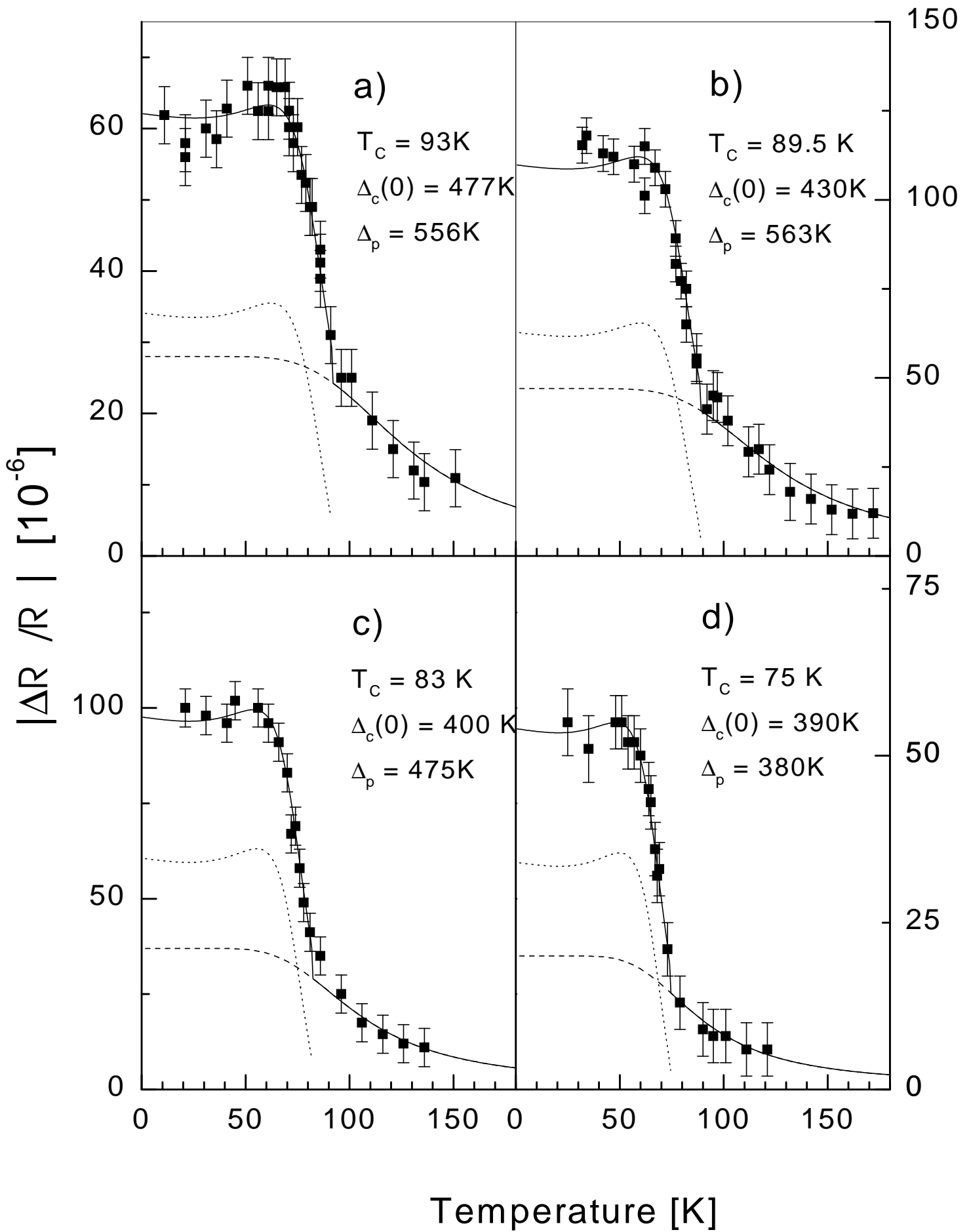
Figure 3. The PI reflection amplitude, $\left| {\sl \Delta }{\cal R}/{\cal R}%
\right| ,$ as a function of $T$ for Y$_{1-x}$Ca$_{x}$Ba$_{2}$Cu$_{3}$O$%
_{7-\delta }$ with a) $x$=0, b) $x$=0.016, c) $x$=0.101 and d) $x$=0.132.
The fits are made using the sum of Eqs. 2 and 3. The values of $\Delta
_{c}(0)$ and $\Delta _{p}$ used in the fit are shown. The separate $A(T)$
and $B(T)$ are also shown dotted and dashed respectively.

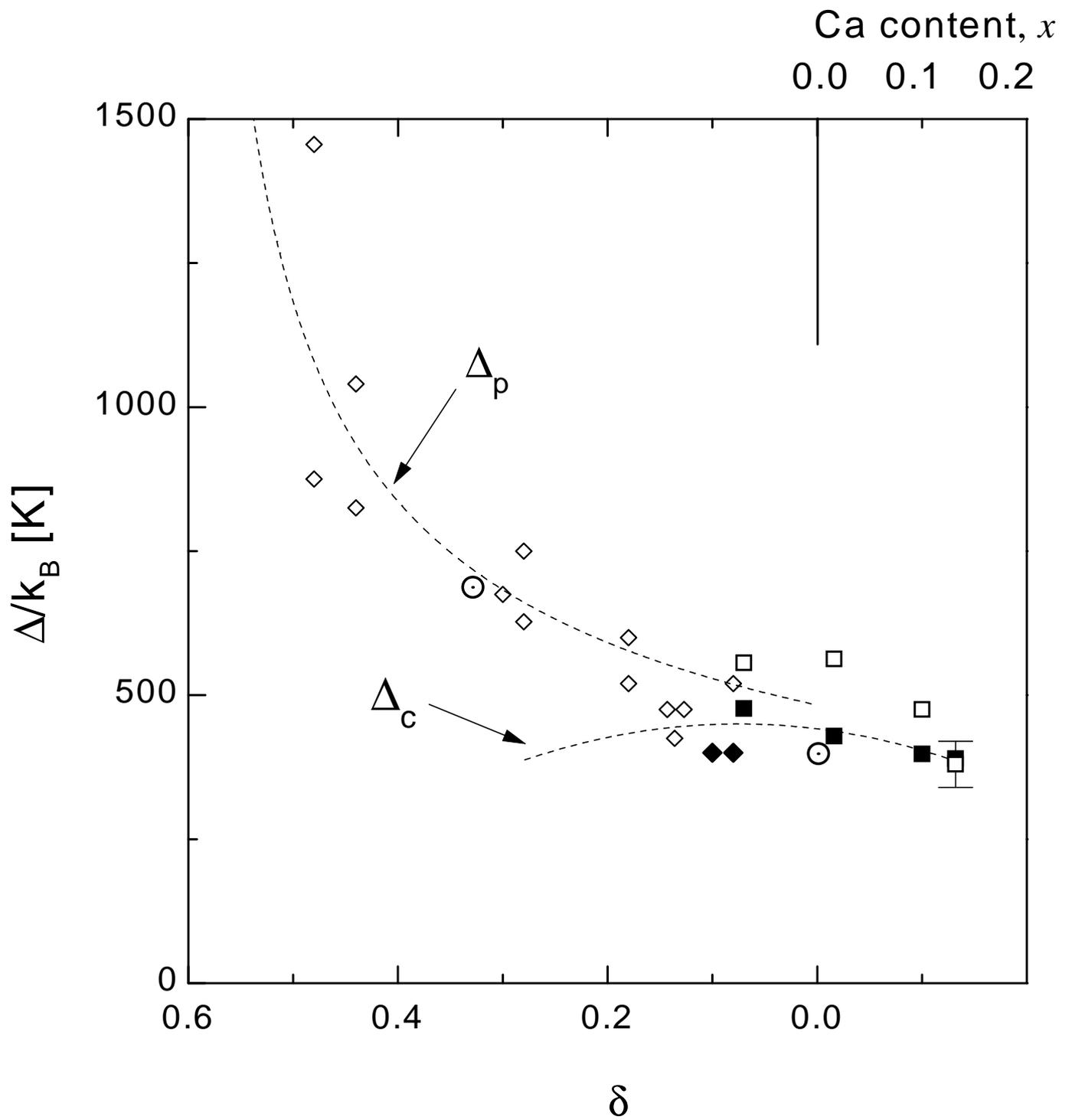
Figure 4. The energy gaps $\Delta _{p}$ and $\Delta _{c}(0)$ as a function
of doping in Y$_{1-x}$Ca$_{x}$Ba$_{2}$Cu$_{3}$O$_{7-\delta }$ obtained from
fits to the data in Figure 3. The open squares represent $\Delta _{p}$,
while the solid symbols are for $\Delta _{c}(0)$ obtained from the fits in
Fig. 3. The open and filled diamonds represent the $\Delta _{p}$ and $\Delta
_{c}(0)$ respectively from Kabanov et al.\cite{Kabanov}. The upper dashed
line represents $\Delta _{p}\propto 1/x,$where $x$ is the carrier
concentration. The lower dashed line is a guide to the eye emphasizing the
behaviour of $\Delta _{c}(0)$. The circles are from tunneling data \cite
{Deutscher}.


\begin{references}
\bibitem{remark1}  In addition, a much slower component is also observed,
similar as in ref.\cite{Stevens}, but this will not be disussed further in
this paper.
\end{references}
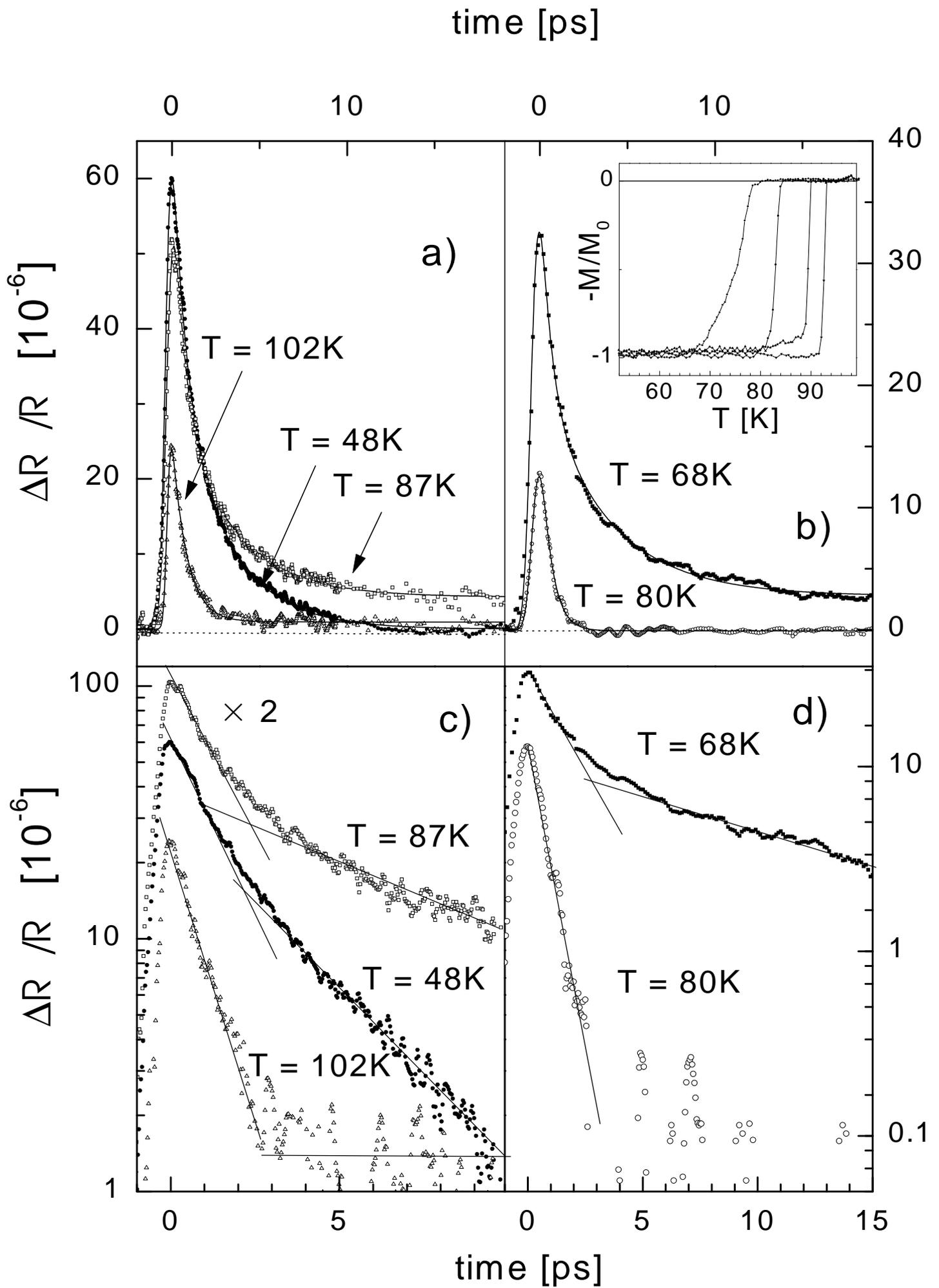
\end{document}